\documentclass[nofootinbib,superscriptaddress,showpacs,showkeys]{revtex4}

\usepackage{graphicx,epsfig}
\usepackage{amsfonts,amsmath,amssymb,amsthm,amscd}
\usepackage[T2A]{fontenc}
\usepackage{color}

\begin{document}

\title{Optimized laser pulse profile for efficient radiation pressure acceleration of ions.}

\author{S. S. Bulanov}
\affiliation{University of California, Berkeley, California 94720, USA}

\author{C. B. Schroeder}
\affiliation{Lawrence Berkeley National Laboratory, Berkeley, California 94720, USA}

\author{E. Esarey}
\affiliation{Lawrence Berkeley National Laboratory, Berkeley, California 94720, USA}

\author{W. P. Leemans}
\affiliation{University of California, Berkeley, California 94720, USA}
\affiliation{Lawrence Berkeley National Laboratory, Berkeley, California 94720, USA}

\begin{abstract}
The radiation pressure acceleration regime of laser ion acceleration requires high intensity laser pulses to function efficiently. Moreover the foil should be opaque for incident radiation during the interaction to ensure maximum momentum transfer from the pulse to the foil, which requires proper matching of the target to the laser pulse. However, in the ultrarelativistic regime, this leads to large acceleration distances, over which the high laser intensity for a Gaussian laser pulse must be maintained. It is shown that proper tailoring of the laser pulse profile can significantly reduce the acceleration distance, leading to a compact laser ion accelerator, requiring less energy to operate.         
\end{abstract}

\maketitle

\section{Introduction}

{\noindent}With the rapid development of laser technology, the interaction of high intensity laser pulses with matter has become a major focus of research both theoretically and experimentally. This interaction can lead to a compact source of high energy electrons, ions, and high frequency radiation with a broad range of applications. In particular the laser driven acceleration of ions has attracted a lot of attention recently due to the fact that these ions may be used for fast ignition \cite{FI}, hadron therapy \cite{medical}, injectors for conventional accelerators \cite{injection}, and radiography of dense targets \cite{radiography}. 

One of the most efficient regimes of laser ion acceleration is the radiation pressure acceleration (RPA, which is also often referred to in the literature as "laser piston" and "light sail") regime \cite{RPA}. This regime is based on the receding relativistic mirror concept, \textit{i.e.} when the laser pulse is reflected by the co-moving mirror its frequency is downshifted by $4\gamma^2$, where $\gamma$ is the Lorentz factor of the mirror. Thus, in the case of relativistic motion, almost all the laser pulse energy can be transferred to the mirror, $~(1-1/4\gamma^2)\mathcal{E}_L$, where $\mathcal{E}_L$ is the incident laser pulse energy. The idea goes back to the papers by Lebedev and Eddington \cite{RPA old}, Veksler \cite{RPA Veksler} and has a close analogy, which was emphasized in many papers on the subject, to the "light sail" scheme for spacecraft propulsion \cite{RPA light sail}. If a solid density foil is utilized as such a mirror, then under the action of the laser pulse radiation pressure the foil will be accelerated to an energy proportional to the laser pulse energy, which makes this scheme of acceleration highly efficient. For the ultra-high laser intensities, radiation friction effects may become important, which was studied in Refs. \cite{RPA RR}. Also in the case of the nonrelativistic motion it was shown that this regime provides an efficient mechanism of laser ion acceleration \cite{RPA low intensity}. There are experimental results that indicate the onset of the RPA regime in laser driven ion acceleration \cite{RPA exp}. 

Usually this acceleration mechanism is considered theoretically assuming total reflection of the laser pulse, \textit{i.e.}, the interaction conditions are such to ensure the maximum efficiency. However the inclusion of the finite reflectivity of the foil changes the results \cite{RT in RPA,unlimited,LightSail}, and moreover enables one to identify the optimal thickness of the foil, which maximizes the energy of ions. In addition to laser ion acceleration \cite{RPA, CE, DCE}, several regimes of laser-foil interaction, including the RPA, can be used to produce relativistic mirrors. These mirrors were first introduced in the laser-gas target interactions, where it was shown that a breaking plasma wake wave is able to reflect the counterpropagating radiation in a form of intense high frequency electromagnetic pulse \cite{LI}. In the case of a laser-foil interaction these mirrors are also able to generate high-frequency intense radiation by reflecting the counterpropagating laser pulses \cite{RM_ion}. In the latter case the properties of the relativistic mirror would also depend on the foil reflectivity. 

In this paper we study the optimization of the RPA regime of laser ion acceleration, based on the analysis of the electromagnetic wave reflection by the thin foil. It is well known that in order to ensure high efficiency of acceleration the foil should be opaque to the laser radiation. However, opaque foils should either have high density or be rather thick, or both. This would increase the number of ions in the irradiated spot, thus decreasing the energy that the laser can transfer per ion. Therefore the most efficient acceleration should happen at the threshold of the foil transparency/opacity, which is governed by the reflection coefficient. At this threshold the foil is opaque for radiation, but this opaqueness is ensured by the minimum possible number of ions. Moreover, for relativistic energies of the foil, the effect of relativistic opacity increases the effectiveness of acceleration. As the foil is accelerated to relativistic energies it becomes less transparent to radiation, \textit{i.e.}, relativistically opaque. The notion of the relativistic opacity is analogous to the notion of the relativistic transparency \cite{RT}, which is widely used in studying ion acceleration from thin foils, where the increasing laser vector-potential (quiver motion of the electrons) makes the foil relativistically transparent (for example, see \cite{RT exp}). We show that by utilizing a laser pulse with a properly tailored intensity profile it is possible to maintain the optimal acceleration conditions during the entire interaction, \textit{i.e.}, the accelerated foil will be at the threshold of opacity/transparency for the incident laser pulse at each instant in time. This would lead to a significant reduction of the acceleration time, which will reduce the requirements on the laser pulse, providing a way to a more compact laser ion accelerator. We note that the idea of laser pulse tailoring for reducing the effects of Rayleigh Taylor instability that accompany the RPA of thin foils was first discussed in Refs. \cite{RT in RPA,unlimited}. For example, consider the acceleration of protons by the RPA to the energy of 10 GeV by a 100 fs laser pulse. According to the results of \cite{RPA} (assuming total reflection) the acceleration time is about 10 ps and the acceleration distance is 3 mm. Such long acceleration distance poses a technological challenge, since it would be extremely difficult to have a very high intensity laser system that would provide a Raleigh length of the order of several millimeters. Another way of maintaining high intensity during the acceleration of the foil over the distance of several millimeters would be the utilization of some external guiding structure, which would ensure the laser pulse propagation without diffraction. However, in this case the group velocity of the laser pulse will be limited to the values smaller than the speed of light, and thus the ion energy will be limited to this group velocity \cite{Shock}. Therefore the profiling of the incident laser pulse offers a way to compact laser ion accelerator with relaxed requirements on the total laser pulse energy needed to achieve certain accelerated ion energy.         

It is well known that the motion of the foil under the action of the laser pulse radiation pressure is described by the equations (we set $c=1$ below throughout the paper) \cite{RPA}
\begin{eqnarray} 
\frac{dp}{dt}=\frac{K\left|E_L[t-x(t)]\right|^2}{4\pi n_e l}\frac{(1+p^2)^{1/2}-p}{(1+p^2)^{1/2}+p}, \label{eqn1}\\
\frac{dx}{dt}=\frac{p}{(1+p^2)^{1/2}}, \label{eqn2}
\end{eqnarray}   
where $n_e$ is the electron density in the foil, $l$ is the thickness of the foil, and the foil momentum $p$ is normalized to the ion rest energy $m_i$. The laser pulse electric field is denoted as $E_L[t-x(t)]$, where $x(t)$ is the position of the foil. The parameter $K=2|\rho|^2+|\alpha|^2$comes from the formula for the radiation pressure, which is the sum of reflected, transmitted and incident electromagnetic (EM) wave momentum fluxes: $P=(E^{\prime 2}_L/4\pi)(1+|\rho|^2-|\tau|^2)$, where $\rho$ and $\tau$ are reflection and transmission coefficients respectively. The energy conservation in this case implies that $|\rho|^2+|\tau|^2+|\alpha|^2=1$, where $\alpha$ is the absorption coefficient. By introducing $\alpha$ we include the fact that part of the incident EM wave energy may be absorbed by the foil. Then $P=(E_L^2/4\pi)(\omega^\prime/\omega)^2 K$, where $\omega$ is the laser frequency. Here primed variables correspond to the moving foil reference frame. The reflection coefficient is in general a function of $p(t)$ and of $t$, $x(t)$ through its nonlinear dependence on the laser pulse amplitude, $a(t)$. In deriving Eq. (\ref{eqn1}) it is assumed that the foil acceleration is rather slow such that at each time moment there is an inertial reference frame in which the foil is at rest. The radiation pressure is then obtained from the balance of momentum fluxes of incident, transmitted, and reflected pulses in this frame.

The solution of Eqs. (\ref{eqn1})-(\ref{eqn2}) \cite{RPA} in the case when the reflection coefficient does not depend on the foil momentum is
\begin{equation} \label{momentum}
p=\frac{1}{2}\left(h_0+W_\rho-\frac{1}{h_0+W_\rho}\right),
\end{equation}
where  
\begin{equation} \label{W}
W_\rho=\frac{2\mathcal{F}_\rho}{n_e l},~~~\mbox{and}~~~\mathcal{F}_\rho=\int\limits_{-\infty}^\psi\frac{\left|\rho E_L(\eta)\right|^2}{4\pi}d\eta.
\end{equation}
Here $\psi=t-x(t)$ is the phase, $h_0=p_0+(1+p_0^2)^{1/2}$ is the integration constant, and $p_0$ is the initial momentum of the foil. In this expression $W_\rho$ has the meaning of the reflected EM wave fluence divided by the area density. We should note here that $W_\rho$ is not the fluence of the laser pulse, but the fluence of the laser pulse fraction which will be reflected at the foil and which is determined by the reflection coefficient.  

Note that the reflection coefficient of the foil is not a relativistic invariant, but depends on the momentum of the foil. Even if the foil was partially transparent to the radiation initially, while it is accelerated it becomes less and less transparent for the co-propagating radiation. Thus at some velocity the receding foil becomes opaque for the radiation, providing more and more efficient momentum transfer from the laser pulse to the foil. This is one of the reasons why the RPA regime is so effective in the relativistic case.    

The paper is organized as follows. In section 2 we review the results on the thin foil reflectivity and discuss the dependence of the thin foil reflection coefficient on the laser pulse intensity and the properties of the foil. We study the solutions of the equations governing the RPA regime of ion acceleration with reflection coefficient taken into account, as well as their dependence on the laser pulse intensity and the properties of the foil, in section 3. We also analyze the numerical solutions of these equations. In section 4 we discuss the optimal laser profile, based on the solution of equations of motion. We conclude in section 5.   

\section{Reflectivity of a thin foil} 

{\noindent}Following Ref. \cite{thin foil} in this section we derive expressions for the reflectivity of a thin foil. A thin foil, \textit{i.e.} with the thickness much less then the radiation wavelength, interacting with the EM wave is described by the wave equation for the dimensionless vector-potential $\mathbf{a}(x,t)$ of the EM wave ($a=e A/m_e$, where $m_e$ and $e$ are the electron mass and charge respectively). The initial conditions $\mathbf{a}(x,0)=\mathbf{a}_0(x)$ and $\partial_t\mathbf{a}(x,0)=\partial_t\mathbf{a}_0(x)$ describe the incident EM wave at the foil. It is also assumed that the current is localized in the foil, thus $(\partial_x^2-\partial_t^2)\mathbf{a}(x,t)=4\pi\delta(x)\mathbf{J}(\mathbf{a})$, where $\delta(x)$ is the Dirac delta function. The solution of the wave equation  can be obtained using the d'Alembert formula:
\begin{equation}\label{integral equation}
\mathbf{a}(x,t)=\mathbf{a}_0(x,t)+2\pi\int\limits_0^{t-|x|}\mathbf{J}[\mathbf{a}(0,\tau)]d\tau.
\end{equation}
If we set $x=0$, we obtain an integral equation for the field inside the foil. Differentiating both sides of Eq. (\ref{integral equation}) over $t$ and employing the explicit form of the current, $\mathbf{J}[\mathbf{a}(0,t)]=\epsilon_0 \mathbf{a}(0,t)/2\pi\left(1+\mathbf{a}^2(0,t)\right)^{1/2}$, yields
\begin{equation}\label{vector-potential}
\frac{d\mathbf{a}(0,t)}{dt}+\epsilon_0\frac{\mathbf{a}(0,t)}{\left(1+\mathbf{a}^2(0,t)\right)^{1/2}}=\frac{d\mathbf{a}_0(0,t)}{dt},
\end{equation}
where the dimensionless parameter $\epsilon_0$ is given by 
\begin{equation}
\epsilon_0=\pi \frac{n_el}{n_{cr}\lambda}. 
\end{equation}
Here $n_e$ is the electron density of the foil, $l$ is the thickness of the foil, $n_{cr}=m_e\omega/4\pi e^2$ is the critical plasma density, $\omega$ and $\lambda$ are the laser pulse frequency and wavelength respectively. The parameter $\epsilon_0$, first introduced in \cite{thin foil}, plays an important role in the theoretical description of the EM wave interaction with solid density foils and, in particular, in connection to the ion acceleration. If the incident wave has the form $a_0(x,t)=a_0(x-t)\exp[i(x-t)]$ and is circularly polarized, then we can search for a solution in a form 
\begin{equation} \label{circular}
a_y+i a_z=b(t)\exp(i\theta). 
\end{equation}
If we assume that $b(t)$ and $\theta(t)$ are slowly varying functions of time, then setting $db(t)/dt=0$ and $d\theta(t)/dt=0$ we obtain the approximate solution of Eq. (\ref{vector-potential})
\begin{equation} \label{a_stationary}
b=2^{-1/2}\left\{\left[(a_0^2-\epsilon_0^2-1)^2+4a_0^2\right]^{1/2}+a_0^2-\epsilon_0^2-1)\right\}^{1/2},
\end{equation} 
\begin{equation}\label{theta_stationary}
\theta=-\arccos\left(b/a_0\right).
\end{equation}
Rewriting the Eq. (\ref{vector-potential}) in terms of the dimensionless electric field, \textit{i.e.}, $\mathbf{e}=-d\mathbf{a}/dt$, we obtain for the transmitted wave 
\begin{equation}
\mathbf{e}_t(x,t)=\mathbf{e}_0(x,t)+\epsilon_0\frac{\mathbf{a}(0,t-|x|)}{\left(1+\mathbf{a}^2(0,t-|x|)\right)^{1/2}},
\end{equation} 
and, consequently, for the reflected wave we have 
\begin{equation}\label{E_r}
\mathbf{e}_r(x,t)=-\epsilon_0\frac{\mathbf{a}(0,t+|x|)}{\left(1+\mathbf{a}^2(0,t+|x|)\right)^{1/2}}.
\end{equation}  
Thus we obtained the amplitude of the transmitted wave in the case of the EM pulse interaction with a thin foil ($l\ll\lambda$). This expression can be used to calculate the efficiency of the RPA regime of laser ion acceleration, since the reflection of the laser radiation at the accelerating foil is the key effect driving this mechanism. The amplitude of the reflected wave is obtained by substituting the expressions (\ref{circular},\ref{a_stationary},\ref{theta_stationary}) into the equation (\ref{E_r}),  
\begin{equation}\label{a_r}
a_r=\epsilon_0\left\{\frac{\left[(a_0^2-\epsilon_0^2-1)^2+4a_0^2\right]^{1/2}+a_0^2-\epsilon_0^2-1)}{\left[(a_0^2-\epsilon_0^2-1)^2+4a_0^2\right]^{1/2}+a_0^2-\epsilon_0^2+1)}\right\}^{1/2}.
\end{equation}
In the two limiting cases of opaque ($a_0\ll \epsilon_0$) and transparent ($a_0\gg\epsilon_0$) foil we find 
\begin{equation}
a_r=\left\{
\begin{tabular}{l}
$\displaystyle\frac{\epsilon_0a_0}{(1+\epsilon_0^2)^{1/2}},~~~a_0\ll min\left[1,\epsilon_0\right] $\\ \\
$\displaystyle\epsilon_0-\frac{\epsilon_0}{2a_0^2},~~~a_0\gg max\left[1,\epsilon_0\right].$
\end{tabular}
\right.
\end{equation} 
Analyzing the form of the reflected wave amplitude given by Eq. (\ref{a_r}) we obtain that the threshold of opacity/transparency corresponds to the condition $a_0\approx\epsilon_0$. This result is connected with the fact that the current in the foil is limited by a finite number of electrons in the foil with velocity $<c$. The regime when almost all the incident radiation is reflected by the foil starts at $a_0\approx \epsilon_0$. In this case the reflected wave amplitude is
\begin{equation}\label{epsilon=a0}
a_r(\epsilon=a_0)= a_0\left[\frac{(1+4a_0^2)^{1/2}-1}{(1+4a_0^2)^{1/2}+1}\right]^{1/2},
\end{equation}  
which for $a_0\ll 1$ gives $a_r\approx a_0^2$, and for $a_0\gg 1$ gives $a_r\approx a_0$. In Fig. 1 we show the dependence of the reflection coefficient, $\rho=a_r/a_0$, on $a_0$. For $a_0<\epsilon_0$ it is constant, $\rho\approx 1$, and for $a_0>\epsilon_0$ it decreases as $a_0^{-1}$. This means that as a foil becomes transparent for radiation, there is still a reflected wave, though its amplitude is decreasing as the amplitude of the incident wave goes up. In the case of $\epsilon_0=a_0$, the reflection coefficient quickly rises from zero to almost unity as the amplitude of the incident wave grows (see Fig. 1b).

\begin{figure}[tbp]
\epsfxsize7cm\epsffile{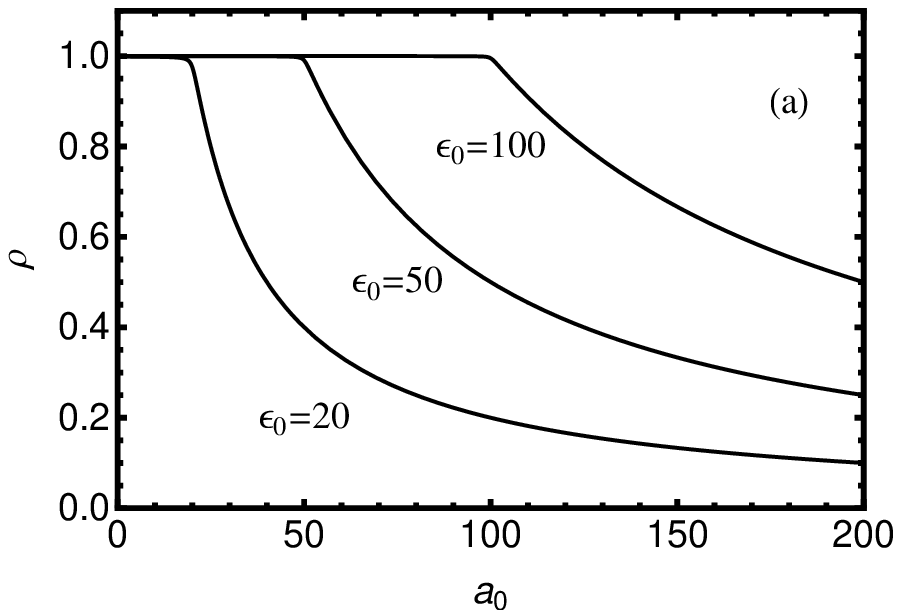} \epsfxsize7cm\epsffile{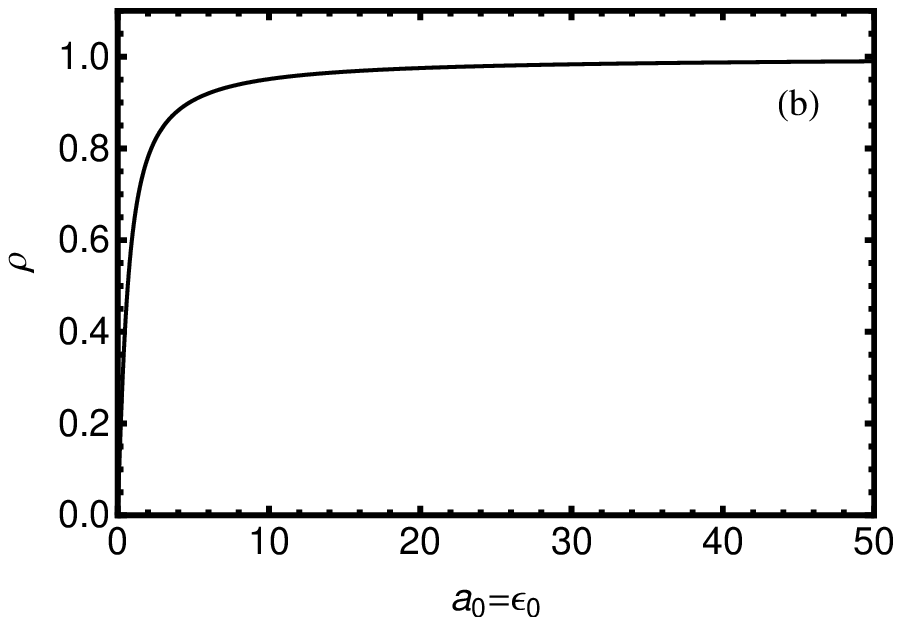} 
\caption{(a) The dependence of the reflection coefficient, $\rho$, on $a_0$ for given $\epsilon=20,~50,~100$; (b) The dependence of the reflection coefficient, $\rho$, on $a_0$ for given $\epsilon=a_0$.}
\end{figure}

\section{Matching the target and the laser pulse for optimal acceleration}

{\noindent} In the previous section we considered the interaction of a foil with an EM pulse in the frame of reference where the foil is at rest. In what follows we are interested in the interaction of a receding foil with the laser pulse. In this case in order to be able to use the formulae obtained in section II we should make a substitution $\epsilon_0\rightarrow\gamma\epsilon_0$, where $\gamma$ is the Lorentz factor of the foil. Thus as the foil is accelerated, the parameter $\epsilon_0$ grows, making the foil opaque for radiation when its velocity reaches some threshold value.

Let us consider two limiting cases ($a_0\gg\gamma\epsilon_0$ and $a_0\ll\gamma\epsilon_0$) of laser pulse interaction with a thin foil in the RPA regime. In these cases the equation of motion can be written in the following form:
\begin{equation} \label{RPA_1}
\frac{dp}{dt}=\frac{m_e}{m_i}\epsilon_0\gamma^2\frac{\gamma-p}{\gamma+p},~~~a_0\gg\gamma\epsilon_0
\end{equation}
\begin{equation} \label{RPA_2}
\frac{dp}{dt}=\frac{m_e}{m_i}\frac{a^2}{\epsilon_0}\frac{\gamma-p}{\gamma+p},~~~a_0\ll\gamma\epsilon_0.
\end{equation}
Equation (\ref{RPA_1}) indicates that in this limit the final energy of ions does not depend on the laser pulse amplitude, but depends on the properties of the target. It is connected with the mentioned above fact that the magnitude of the current in the foil is limited by the total number of electrons in the irradiated spot, and thus the amplitude of the reflected pulse is also limited. However as $\epsilon_0$ increases so does the current in the foil, which leads to the increase of the reflected pulse amplitude and subsequently to the increase of accelerated ion energy. It can also be seen from the solution of equation (\ref{RPA_1}), which in the limit $(m_e/m_i)\epsilon_0\tau\ll 1$ has the form $p\sim (m_e/m_i)\epsilon_0\tau$, where $\tau$ is the duration of the laser pulse in units of wave period. The solution of (\ref{RPA_2}) is given by (\ref{momentum}) and (\ref{W}), where $\rho=1$ and $W_\rho$ is rewritten in the following form: $W_\rho=\int_{-\infty}^{\tau}(m_e/m_i)(a(\eta)^2/\epsilon_0)d\eta$ to show the dependence on $\epsilon_0$. Thus we have two limiting cases of $a_0\gg\gamma\epsilon_0$ and $a_0\ll\gamma\epsilon_0$, in the first case the energy of ions grows with the increase of $\epsilon_0$, while in the second case it goes down with the increase of $\epsilon_0$. This means that there should be a maximum of ion energy for a given value of $a_0$, i.e. for each value of the laser pulse amplitude there is an optimal thickness of the target, which maximizes the accelerated ion energy.       

\begin{figure}[tbp] 
\epsfxsize5cm\epsffile{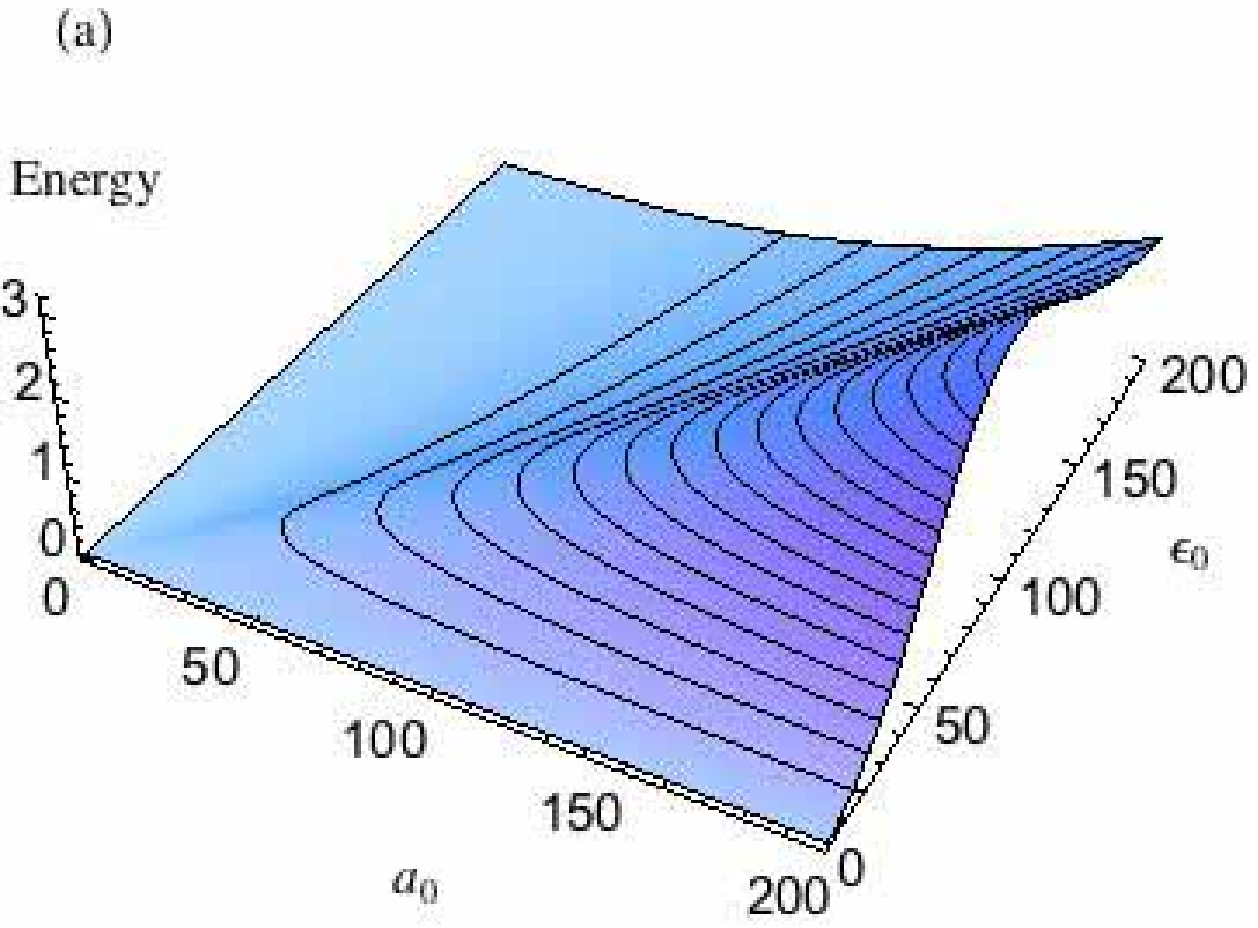} \epsfxsize5cm\epsffile{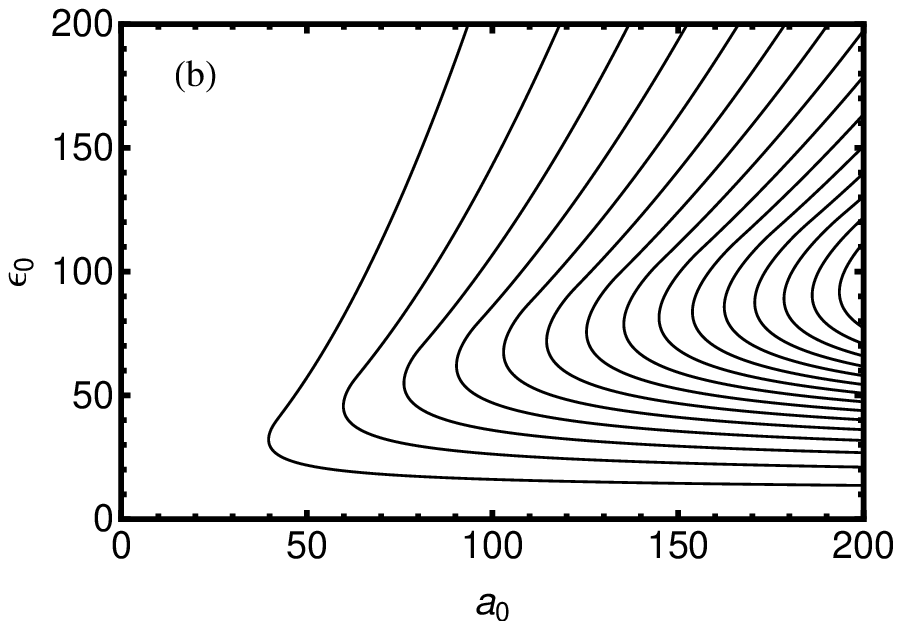} \epsfxsize5cm\epsffile{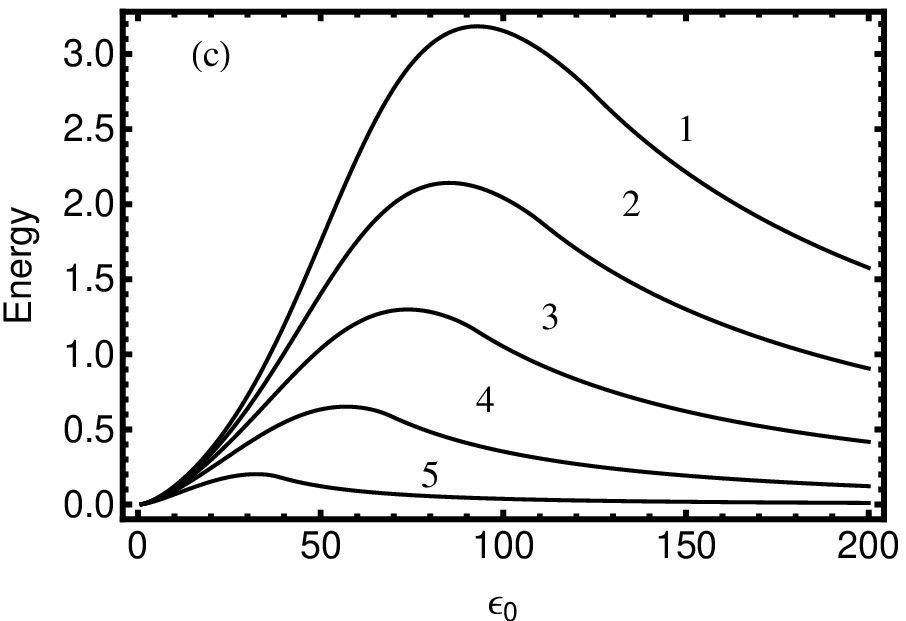}
\epsfxsize5cm\epsffile{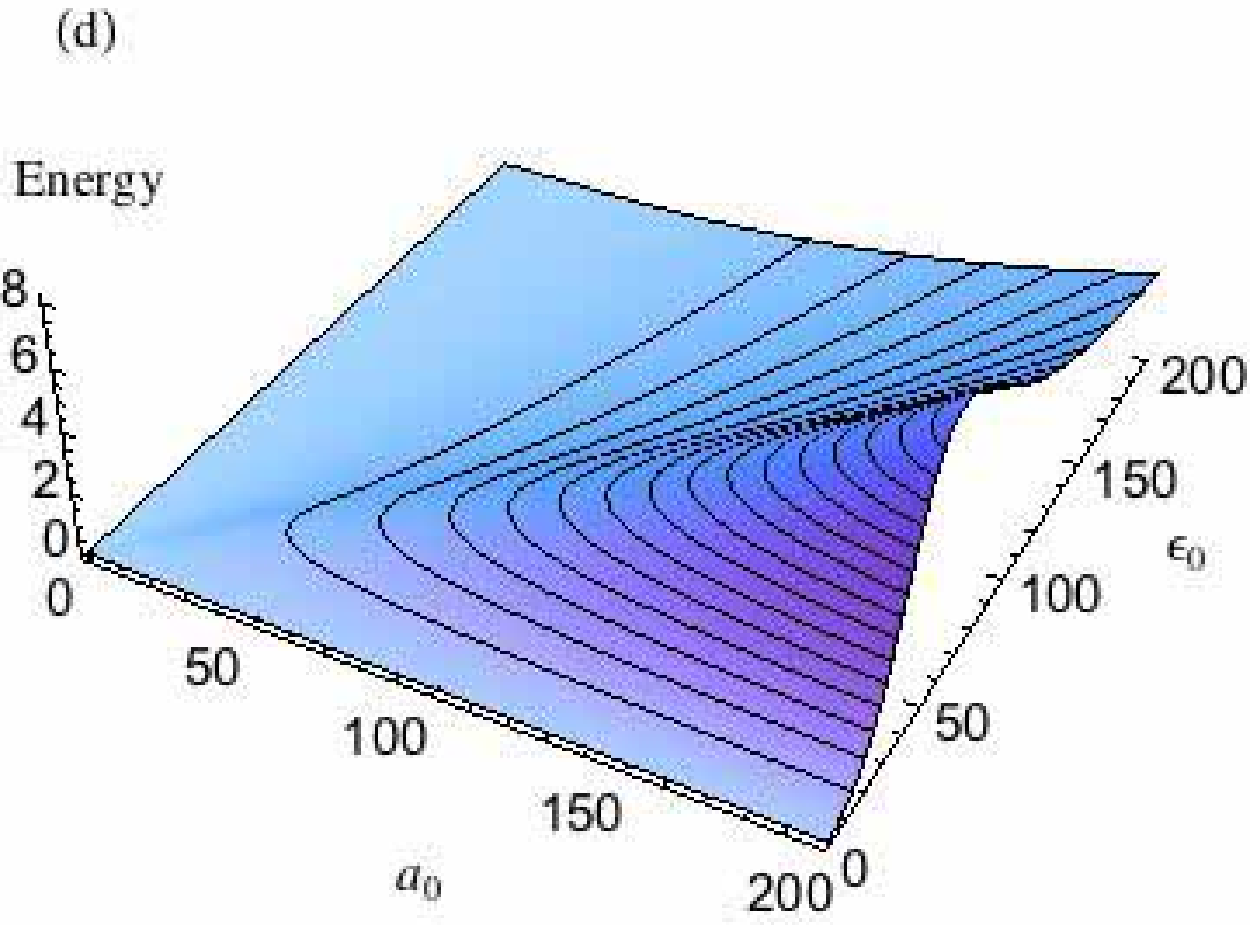} \epsfxsize5cm\epsffile{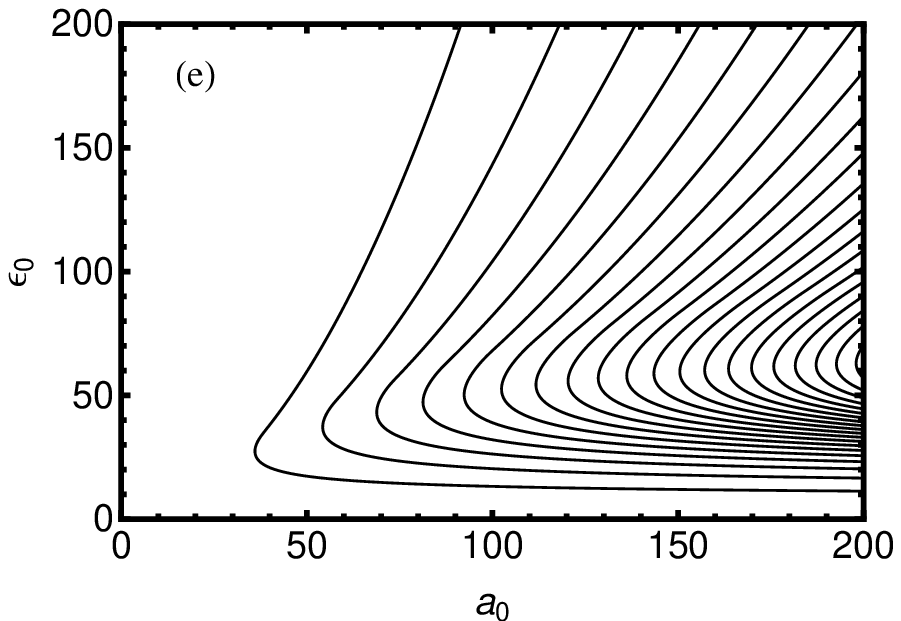} \epsfxsize5cm\epsffile{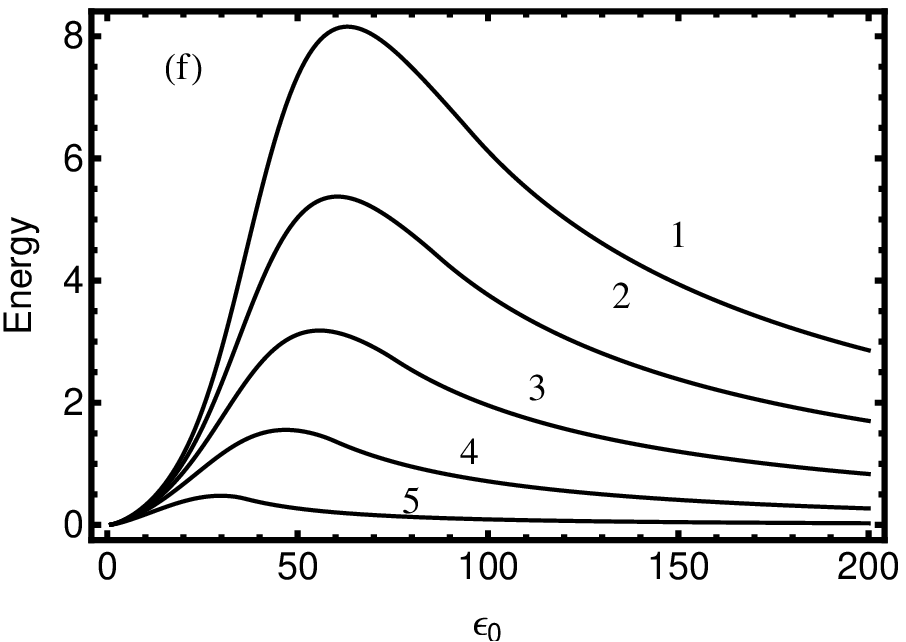}
\epsfxsize5cm\epsffile{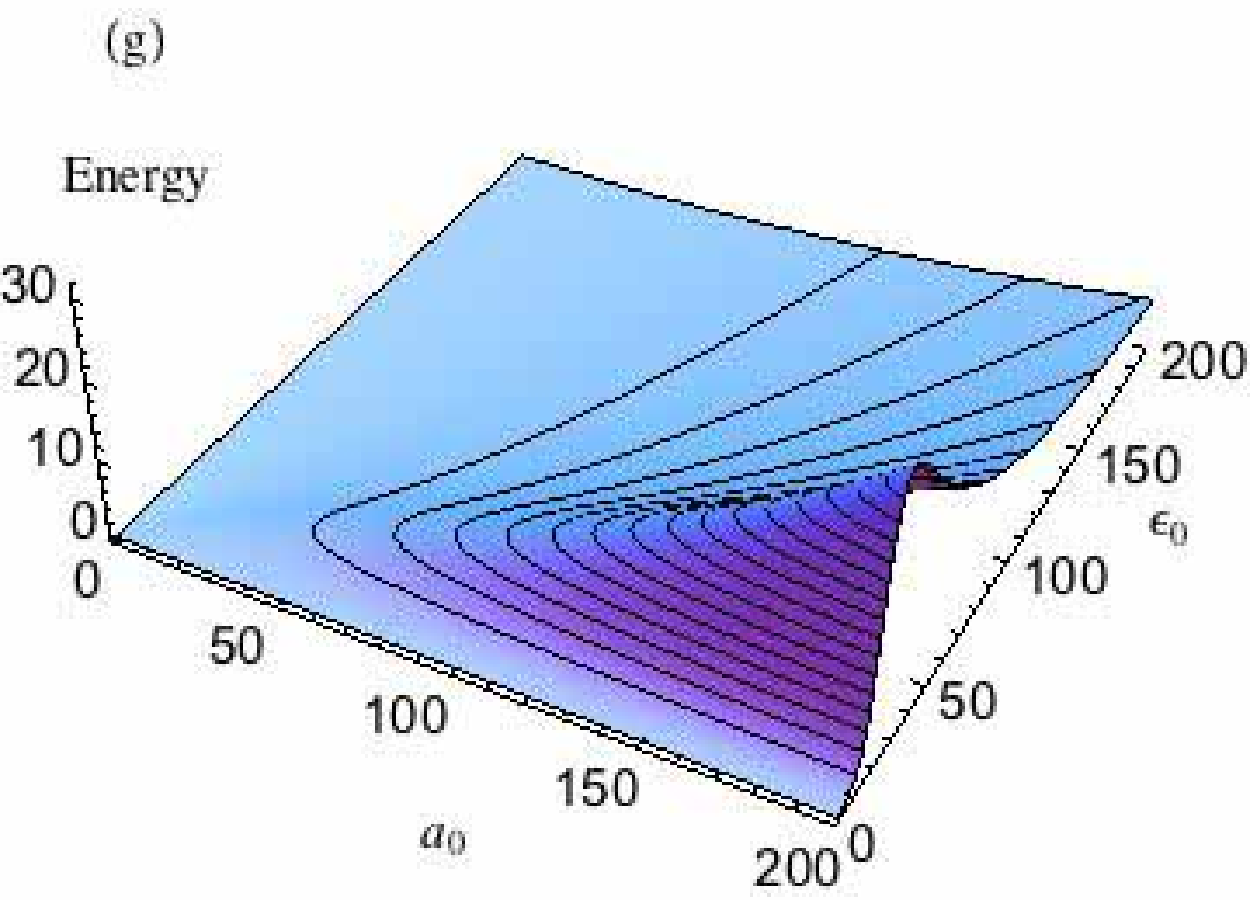} \epsfxsize5cm\epsffile{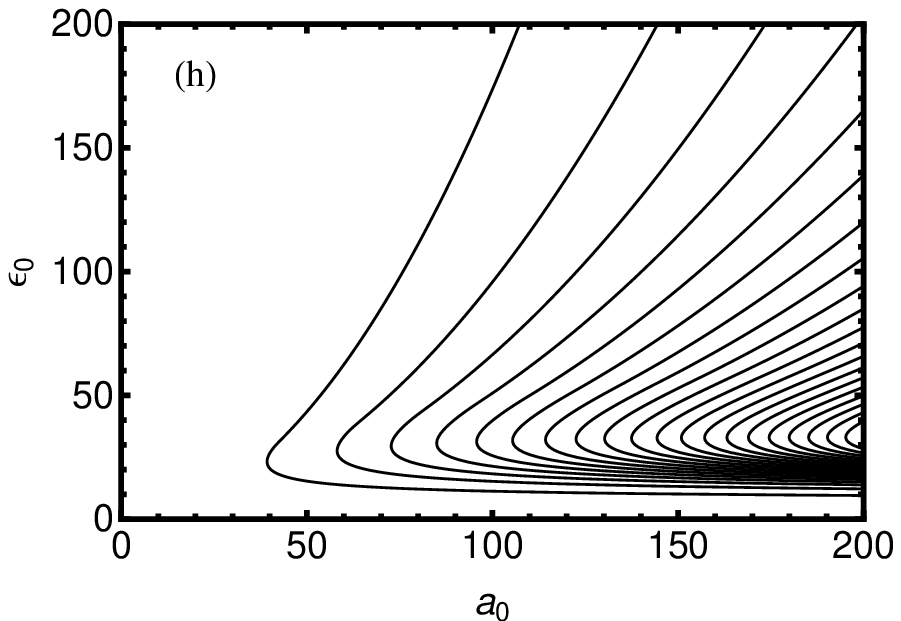} \epsfxsize5cm\epsffile{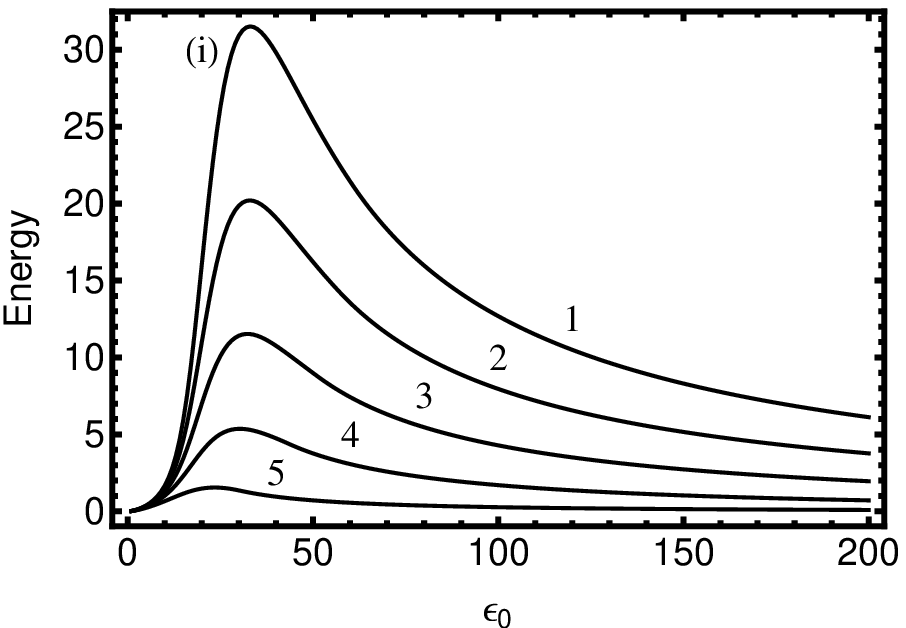}
\caption{\label{curves} The distribution of the maximum ion kinetic energy in the ($a_0,\epsilon_0$) plane for different laser pulse durations: 30 fs (first row), 50 fs (second row), and 100 fs (third row). The 3D plots of the distribution are shown in (a,d,g), contour plots are shown in (b,e,h). The contours correspond to the following values of the ion energy: 30 fs pulse duration - (0.2, 0.4, 0.6,...,3.0), 50 fs pulse duration - (0.4, 0.8, 1.2,...,8.0), and 100 fs pulse duration - (1.5, 3.0, 4.5,...,30.0). The lineouts of the distribution at $a_0=$ 200 (1), 160 (2), 120 (3), 80 (4), and 40 (5) are shown in (c,f,i). The ion energy is measured in the units of ion rest energy.}
\end{figure}

\begin{figure}[tbp]
\epsfxsize7cm\epsffile{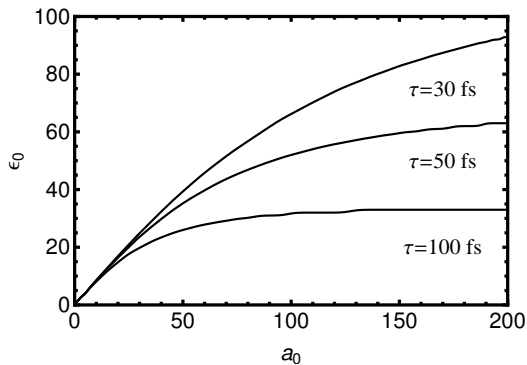}  
\caption{\label{matching} The curves in the ($a_0,\epsilon_0$) plane, along which the maximum ion energy is achieved.}
\end{figure}

In order to find this maximum we solve the equations of motion (\ref{eqn1}, \ref{eqn2}) numerically. We consider RPA by a finite duration laser pulse. The longitudinal profile of the pulse is chosen to be Gaussian. The vector-potential has the form $a(\psi)=a_0 \exp(-4\psi^2/\tau^2)$, the laser wavelength is $\lambda=0.8~\mu$m, and the ions are chosen to be protons $m_i=m_p$. The foil is characterized by the parameter $\epsilon_0$, which was varied from 1 to 200. The amplitude of the laser pulse vector-potential was varied from 1 to 200, several laser pulse durations were considered, from 30 fs to 100 fs. The results of the numerical solution of the equations of motion are shown in Fig. \ref{curves}. We varied the vector-potential of the pulse and the parameter $\epsilon_0$ of the foil. For each pair $(a_0,\epsilon_0)$ we solved the equations of motion, obtaining the final energy of the laser accelerated thin foil. In Fig. \ref{curves} in the first two columns the distributions of the final ion energy in the $(a_0,\epsilon_0)$ plane are shown for the pulse durations of 30 fs, 50 fs, and 100 fs. These distributions indicate the existence of an optimal value of the parameter $\epsilon_0$, which maximizes the energy of ions, for each value of $a_0$. It can be seen more clearly from the third column in Fig. \ref{curves}, where the ion energy dependences on parameter $\epsilon_0$ for several fixed values of $a_0$ are shown. It can be seen from the contour plots that for small values of $\epsilon_0$ the contours of equal energy are almost horizontal, \textit{i.e.} the accelerated ion energy does not depend on the laser pulse amplitude and is determined by the value of parameter $\epsilon_0$. It corresponds to the limit $a_0\gg\gamma\epsilon_0$, described by Eq. (\ref{RPA_1}). Whereas for large values of $\epsilon_0$ the ion energy increases with the increase of $a_0$, which corresponds to the limit $a_0\ll\gamma\epsilon_0$, described by Eq. (\ref{RPA_2}). If, using the results shown in Fig. \ref{curves}, we plot the points $(a_0,\epsilon_0)$ corresponding to the maximum ion energy for fixed $a_0$, then for each laser pulse duration we will get the curve $\epsilon_0=\epsilon_0(a_0)$. These curves are shown in Fig. \ref{matching}. It can be seen that for small $a_0$ the curves approximately follow an $\epsilon_0=a_0$ line. The smaller is the energy in the pulse (\textit{i.e.}, the smaller is the duration of the pulse) the longer the curve follow the $\epsilon_0=a_0$ line. This is connected with the fact that for low energy laser pulses the ions are accelerated by RPA to nonrelativistic energies. Such dependence for the optimal laser ion acceleration from thin foils was reported in Ref. \cite{optimal_esirkepov}. As the laser pulse duration is increased and more energy is transferred from the laser pulse to the foil the curve $\epsilon_0=\epsilon_0(a_0)$ begins to deviate from the $\epsilon_0=a_0$ line. Thus the optimal acceleration is achieved at such values of $a_0$ and $\epsilon_0$ that $\epsilon_0<a_0$. Also the ions for such values of $\epsilon_0$ and $a_0$ are accelerated to relativistic energies. Such behavior of the interaction indicates that the accelerated foil is opaque for incident radiation, which ensures almost total reflection and consequently the maximum energy transfer from the pulse to the foil. Thus the regime of relativistic opacity manifests itself in laser ion acceleration from thin foils in the RPA regime. The curves $\epsilon_0=\epsilon_0(a_0)$, shown in Fig. \ref{matching}, indicate how the amplitude of the laser pulse should be matched to the parameter $\epsilon_0$ of the foil to ensure optimal conditions for laser driven ion acceleration.  

\section{Laser pulse profiling for optimal acceleration}

It is plausible to assume that the condition of the laser pulse matching to the foil should be in the lab frame of the form $a_0=\gamma\epsilon_0$, which corresponds to the threshold of the moving foil opacity/transparency. Such condition will ensure the maximum momentum transfer from the laser to the foil, since the foil is opaque for radiation, also the opacity will be maintained by the minimum possible number of ions, thus increasing the energy per ion value. However in the case of the laser pulse with gaussian longitudinal profile, considered in the previous section, such a condition can not be maintained during the entire acceleration process. In what follows we solve the equations of motion of a foil accelerated by the laser pulse radiation pressure in order to find the laser pulse profile which would ensure that the condition $a=\gamma\epsilon_0$ is maintained during the acceleration process. We rewrite the Eq. (\ref{eqn1}) in the following form, taking into account that the condition $a=\gamma\epsilon_0$ should be maintained at every time instant of the acceleration process,
\begin{equation}\label{gamma_epsilon=a0}
\frac{dp}{d(\omega t)}=\epsilon_0\frac{m_e}{m_i}\rho^2\gamma^2\frac{\gamma-p}{\gamma+p},
\end{equation}   
where $\omega$ is the frequency of the laser pulse, and the reflection coefficient, from Eq. (\ref{epsilon=a0}) with $a=\gamma\epsilon_0$ is
\begin{equation}
\rho=\left[\frac{(1+4\gamma^2\epsilon_0^2)^{1/2}-1}{(1+4\gamma^2\epsilon_0^2)^{1/2}+1}\right]^{1/2}.
\end{equation}
The solution of the equation of motion, Eq. (\ref{gamma_epsilon=a0}), can be written in quadratures:
\begin{equation}\label{quad}
\epsilon_0\frac{m_e}{m_i}\omega t=F(p),
\end{equation}
where 
\begin{equation}
F(p)=\int_0^p dp^\prime\rho^{-2}\gamma^{-2}(\gamma^\prime+p^\prime)/(\gamma^\prime-p^\prime).
\end{equation} 
In the limit $t\rightarrow\infty$ the solution to Eq. (\ref{quad}) is $p\sim t$. Such time dependence is in striking difference with a case of constant amplitude laser pulse, considered in Ref. \cite{RPA}, where $p\sim t^{1/3}$. Hence the acceleration time for the profiled pulse should be less than in the case of a uniform profile.

Equation (\ref{gamma_epsilon=a0}) can be solved numerically. The results of the numerical solution are shown in Fig. 4 for different values of the parameter $\epsilon_0$. As in the previous section we choose the ions to be protons, $m_i=m_p$. In Fig. 4a one can see the dependence of the accelerated ion energy on time. As it could be expected from Eq. (\ref{quad}) the dependence is linear. In Fig. 4b the temporal profiles of the laser pulse, corresponding to the condition $a=\gamma\epsilon_0$, are shown. All the curves demonstrate similar behavior. They have a singular point at some $t=\psi_*$, which is determined by the parameter $\epsilon_0$. In order to compare a profiled pulse case with a pulse having a constant field strength, $E_L=constant$, we present Fig. 4c, where the dependence of the ion energy on time is shown for these two cases. It takes several orders of magnitude more time to reach the same energy for the constant pulse then for a profiled one. Thus the utilization of profiled pulses will reduce the acceleration time.     

\begin{figure}[tbp]
\epsfxsize5cm\epsffile{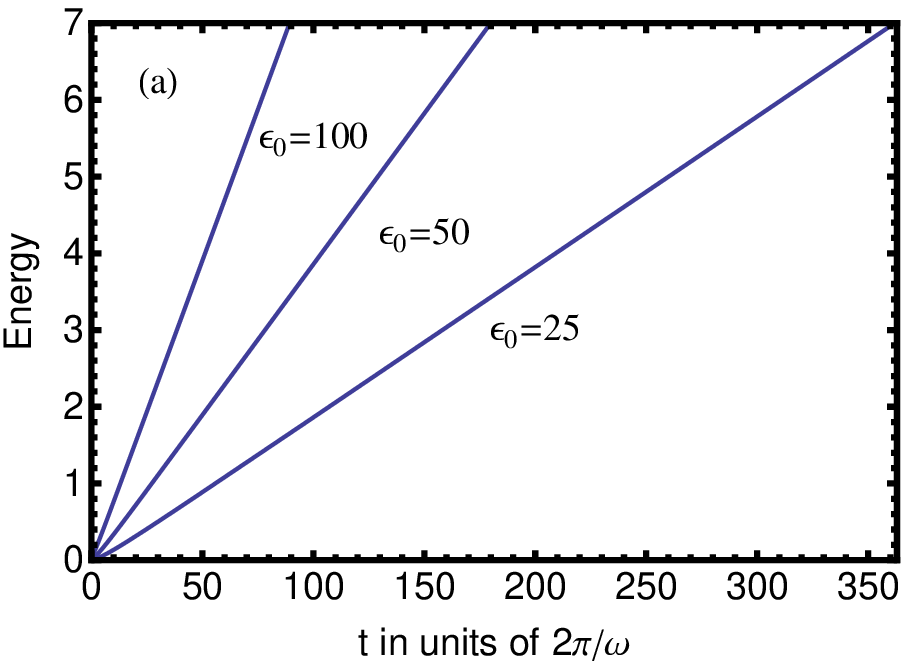} \epsfxsize5cm\epsffile{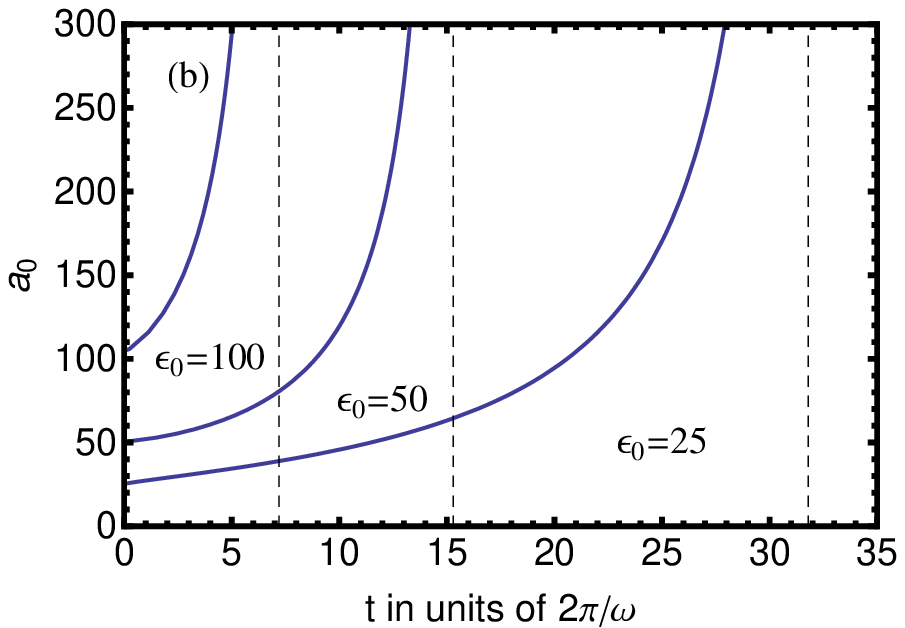} \epsfxsize5cm\epsffile{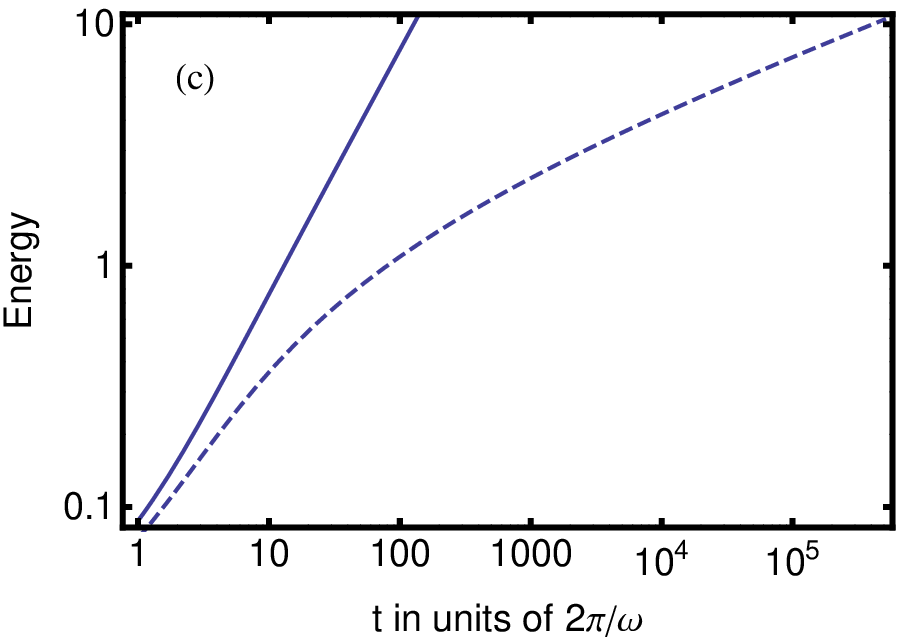} 
\caption{(a) The dependence of the accelerated ion energy on time for $\epsilon_0=25,~50,~100$ from bottom to top; (b) The profiles of the laser pulses for $\epsilon_0=25,~50,~100$, which are able to maintain the condition $a=\gamma\epsilon_0$ during the acceleration process, from bottom to top. Dotted vertical lines denote the position of $\psi_*$ for each of the three values of $\epsilon_0$: $\psi_*(\epsilon_0=25)=31.8$, $\psi_*(\epsilon_0=50)=15.3$, and $\psi_*(\epsilon_0=100)=7.2$; (c) The dependence of the accelerated ion energy on time for a profiled laser pulse (solid line) and for a pulse with $E_L=constant$ (dashed line) for $\epsilon_0=50$. The ion energy is measured in units of ion rest energy.}
\end{figure}

In order to study the behavior of the profiled laser pulse, let us consider a limiting case of ultrarelativistic ion energies. In this case $p\gg 1$, $\rho=1$, and 
\begin{equation}
p=\frac{1}{2}\int\limits_{-\infty}^\psi \frac{|\rho E|^2 d\eta}{2\pi n_e l}. 
\end{equation}
If we rewrite the condition $a=\gamma\epsilon_0$ in the form of an equation       
\begin{equation}
a=\frac{1}{4}\int\limits_{-\infty}^\psi\frac{E^2 d\eta}{n_{cr}\lambda m_i},
\end{equation}
then differentiating both sides of this equation with respect to the wave phase $\psi$ and expressing $a$ in terms of the electric field, we obtain the following differential equation:
\begin{equation}\label{profile}
\frac{dE}{d\psi}=\frac{eE^2}{2m_i}.
\end{equation} 
The solution of (\ref{profile}) is
\begin{equation}\label{field}
E=\frac{E_0}{1-\psi/\psi_*},
\end{equation}
where $E_0=2\pi e n_e l$ is the initial field amplitude, determined by the properties of the foil, and 
\begin{equation}\label{psi*}
\psi_*=\frac{2m_i}{eE_0}=\frac{m_i}{m_e}\frac{2}{\omega a_0}
\end{equation} 
is the maximum duration of the pulse, since at $\psi=\psi_*$ the electric field goes to infinity. The solution, Eq. (\ref{field}), demonstrates the same behavior as the numerical solution of Eq. (\ref{gamma_epsilon=a0}), \textit{i.e.} the singularity at $\psi=\psi_*$. The analytical estimate of the value of $\psi_*$ is in reasonable agreement with the result of the numerical solution of the equation of motion, Eq. (\ref{gamma_epsilon=a0}). This form of the pulse is a special case of a profile considered in Ref. \cite{RT in RPA}, $E\sim (1-\psi/\psi_m)^{-\alpha}$, where $\alpha$ and $\psi_m$ are free parameters. Such a profile, according to the results of Ref. \cite{RT in RPA} suppresses the development of the Rayleigh Taylor (RT) instability that can tear the foil apart transversely and effectively stop the acceleration. Thus the profile (\ref{field},\ref{psi*}) not only optimizes the acceleration but also suppresses the RT instability. 

We note that the optimal profile (\ref{field},\ref{psi*}) is stable in a sense that for a profile very close to it, $E\sim [1-\psi/(\psi_*+d\psi_*)]^{-(1+d\alpha)}$, where $d\psi_*\ll 1$ and $d\alpha\ll 1$, the condition $a(t)=\gamma(t)\epsilon_0$ is maintained with the accuracy up to the terms linear in $d\psi_*$ and $d\alpha$, which increase logarithmically with time. 

If we consider the case of a profile different from an optimal one (\ref{field},\ref{psi*}), $E\sim (1-\psi/\psi_m)^{-\alpha}$, where $\psi_m\neq \psi_*$ and $\alpha\neq 1$, the condition $a(t)=\gamma(t)\epsilon_0$ can not be maintained at each time instant during the interaction even approximately. For $\alpha>1$, $a(t)<\gamma(t)\epsilon_0$, which means that the foil is opaque to radiation, but this opaqueness is not maintained by the minimal possible number of ions. This will reduce the maximum energy of each individual ion. In the case $\alpha>1$, $a(t)>\gamma(t)\epsilon_0$ and the foil at some point will become transparent for radiation, reducing the acceleration efficiency. 

Let us compare the acceleration times in the case of a profiled laser pulse and in the case of a constant amplitude laser pulse. The energy in both cases is set to be the same. The duration of the profiled pulse is denoted as $\tau_{las}$ and of constant amplitude pulse as $T_{las}$. For constant laser energy $T_{las}=\tau_{las}/(1-\tau_{las}/\psi_{*} )$.

We can rewrite the second equation of motion, Eq. (\ref{eqn2}), in terms of the phase $\psi$
\begin{equation}
\frac{d\psi}{dt}=\frac{2}{(h_0+W_\rho)^2+1}.
\end{equation}     
Then in the case of $E_L=constant$ we obtain
\begin{equation}
t=\frac{1}{2}(h_0+1)\psi+\frac{1}{2}h_0\kappa^2\psi^2+\frac{1}{6}\kappa^2\psi^3.
\end{equation}
Here we took $W_\rho$ in the form $W_\rho=\kappa\psi$ with $\kappa=2\mathcal{E}_L^\rho/N_i m_i$, where $\mathcal{E}_L^\rho$ is the energy of the incident laser pulse fraction, which is reflected by the foil, and $N_i$ is the total number of ions in the irradiated spot. For $t\rightarrow\infty$ we obtain for the time of acceleration the following expression: 
\begin{equation}
T_{acc}\approx \frac{1}{6}W_\rho^2\tau^\prime_{las}.
\end{equation}
In the case of a profiled laser pulse the time of acceleration is different:
\begin{equation}
t_{acc}=\frac{W_\rho^2}{2}\frac{\psi_*^2}{\tau_{las}^2}\left(1-\frac{\tau_{las}}{\psi_*}\right)\tau_{las}.
\end{equation}
In order to compare the acceleration times in the two mentioned above cases we calculate their ratio:
\begin{equation}
\frac{T_{acc}}{t_{acc}}=\frac{1}{3}\left(\frac{\tau_{las}/\psi_*}{1-\tau_{las}/\psi_*}\right)^2.
\end{equation}
This ratio can also be written in terms of laser pulse energy:
\begin{equation}
\frac{T_{acc}}{t_{acc}}=\frac{1}{3}\left(\frac{\mathcal{E}_L}{\mathcal{E}_*}\right)^2,
\end{equation}
where $\mathcal{E}_*=(E_0^2/4\pi)\pi R^2\psi_*$ is some characteristic energy, which corresponds to the energy of profiled laser pulse with the duration half of maximum one, $\tau_{las}=\psi_*/2$, or to the energy of a constant amplitude laser pulse with the duration of $\psi_*$. Thus we see that for $\tau_{las}>\psi_*/2$ there is a significant reduction of the acceleration time in the case of a profiled laser pulse. If $\tau_{las}=0.9\psi_*$, then $t_{acc}\approx 10^{-2}T_{acc}$. This means that the utilization of the profiled laser pulses can significantly reduce the length of laser ion accelerators operating in the RPA regime.    

\section{Conclusions}

We studied the behavior of the radiation pressure acceleration (RPA) regime of laser ion acceleration in the case of the laser pulse interaction with ultra-thin foils of solid density. Particular attention was paid to the contribution of the foil reflectivity to the process of acceleration in terms of maximum achievable ion energy and the efficiency of acceleration. The analysis was performed in terms of two parameters: the amplitude of the laser pulse vector-potential, $a_0$, and the parameter $\epsilon_0$ \cite{thin foil}, which governs the transparency/opacity of the foil, \textit{i.e.} for $a_0>\epsilon_0$ the foil is referred to as transparent and for $a_0<\epsilon_0$ the foil is referred to as opaque, assuming that the foil is stationary (nonrelativivstic).   

Based on the analysis of the electromagnetic wave reflection by a thin foil, we studied the matching of the laser pulse to the properties of the foil so that the interaction of the laser pulse with the foil would provide optimal conditions for ion acceleration in the RPA regime. We showed that for small values of $a_0$ and $\epsilon_0$ the accelerated ion energy is maximal when $a_0=\epsilon_0$, which is in agreement with the results of Ref. \cite{optimal_esirkepov}. However as $a_0$ and $\epsilon_0$ increase, the relation between them, which provide maximum accelerated ion energy, begins to deviate from $a_0=\epsilon_0$ to such values of $a_0$ and $\epsilon_0$ that $a_0>\epsilon_0$. This behavior indicates that onset the regime of the relativistic opacity. The foil accelerated to relativistic energies becomes opaque for the co-propagating electromagnetic wave, even if it was transparent when it was at rest. Thus the relativistic opacity provides maximum momentum transfer from the laser pulse to the foil. 

The analysis of the reflection coefficient of a thin foil indicates that the optimal acceleration conditions, which provide maximum ion energy, are ensured when the interaction of a laser pulse with the foil happens at the threshold of the foil opacity/transparency. Or, in other words, the foil is opaque to radiation, but this opacity is provided by minimum possible number of ions. Thus the momentum transfer from the pulse to the foil is maximal and also the energy per ion in the irradiated spot is maximal.  

For the relativistic foil the condition of the laser pulse matching to the foil, or the threshold of opacity/transparency,  is of the form $a_0=\gamma\epsilon_0$. However, in the case of the laser pulse with Gaussian longitudinal profile, this condition can not be maintained during the entire acceleration process. We solved the equations of motion of a foil accelerated by the laser pulse radiation pressure in order to find the laser pulse profile which would ensure that the condition $a=\gamma\epsilon_0$ is maintained during the acceleration process and found that it is of the form $E=E_0/(1-\psi/\psi_*)$, where $\psi=t-x(t)$ is the phase of the pulse and $\psi_*$ is given by the Eq. (\ref{psi*}). Such form of the pulse, according to the results of \cite{RT in RPA} also suppresses the development of the RT instability, which is important for the RPA regime. We showed that the matching of the laser pulse profile leads to significant reduction of the acceleration time and thus acceleration distance. This is a critical parameter for the RPA scheme of laser ion acceleration, since the acceleration distance is of the order of the required Rayleigh length for the  high intensity laser systems. Therefore the incident laser pulse of the form described in this paper offers a reasonable approach to compact laser ion accelerator, with relaxed requirements on the total laser pulse energy needed to achieve certain accelerated ion energy. 

We appreciate support from the NSF under Grant No. PHY-0935197 and the Office of Science of the US DOE under Contract No. DE-AC02-05CH11231 and No. DE-FG02-12ER41798.


\begin{thebibliography}{99}
\bibitem{FI} M. Roth, T. E. Cowan, M. H. Key, S. P. Hatchett, C. Brown,
W. Fountain, J. Johnson, D. M. Pennington, R. A. Snavely, S. C. Wilks, K. Yasuike,
H. Ruhl, F. Pegoraro, S. V. Bulanov, E. M. Campbell, M. D. Perry, and H. Powell,
Phys. Rev. Lett. 86, 436 (2001); V. Yu. Bychenkov, W. Rozmus, A. Maksimchuk,
D. Umstadter and C. E. Capjack, Plasma Phys. Rep. 27, 1017 (2001);
A. Macchi, A. Antonicci, S. Atzeni, D. Batani, F. Califano, F. Cornolti,
J. J. Honrubia, T. V. Lisseikina, F. Pegoraro, and M. Temporal, Nucl. Fusion 43, 362 (2003);
J. J. Honrubia, J. C. Fernandez, M. Temporal, B. M. Hegelich, and J. Meyer-ter-Vehn,
Physics of Plasmas 16, 102701 (2009).
\bibitem{medical} S. V. Bulanov and V. S. Khoroshkov, Plasma. Phys. Rep. \textbf{28}, 453
(2002).
\bibitem{injection} K. Krushelnick, E. L. Clark, R. Allott, F. N. Beg, C. N. Danson, A. Machacek, V. Malka, Z. Najmudin, D. Neely, P. A. Norreys, M. R. Salvati, M. I. K. Santala, M. Tatarakis, I. Watts, M.
Zepf, A. E. Dangor, Plasma Science, IEEE Transactions on 28, 1184 -
1189 (2000).
\bibitem{radiography} M. Borghesi, J. Fuchs, S. V. Bulanov, A. J. Mackinnon, P. K. Patel, and M. Roth, Fusion Science and Technology \textbf{49}, 412
(2006).
\bibitem{RPA} T. Esirkepov, M. Borghesi, S. V. Bulanov, G. Mourou, and T. Tajima, \textit{et al.}, Phys. Rev. Lett. \textbf{92}, 175003 (2004).
\bibitem{RPA old} P. N. Lebedev, Ann. Phys. (Leipzig) \textbf{6}, 433 (1901); A. S. Eddington, Mon. Not. R. Astron. 
Soc. \textbf{85}, 408 (1925).
\bibitem{RPA Veksler} V. I. Veksler, Sov. J. Atomic Energy \textbf{2}, 525 (1957).
\bibitem{RPA light sail} F. A. Zander, Technika i Zhizn, No. \textbf{13}, 15 (1924) [in Russian]; R. L. Forward, Missiles and Rockets \textbf{10}, 26 (1962); G. Marx, Nature \textbf{211}, 22 (1966); J. L. Redding, Nature, 213, 588 (1967).
\bibitem{RPA RR}  N. N. Naumova, T. Schlegel, V. T. Tikhonchuk, C. Labaune, I. V. Sokolov, G. Mourou, Phys. Rev. Lett.  \textbf{102}, 025002 (2009); T. Schlegel, N. Naumova, V. T. Tikhonchuk, C. Labaune, I. V. Sokolov, G. Mourou, Phys. Plasmas \textbf{16}, 083103 (2009); M. Tamburini, F. Pegoraro, A. Di Piazza, C. Keitel, A. Macchi, New J. Phys. \textbf{12}, 123005 (2010); M. Chen, A. Pukhov, T.-P. Yu, Z.-M. Sheng, Plasma Phys. Control. Fusion \textbf{53}, 014004 (2011); U. Sinha and P. Kaw, Phys. Plasmas \textbf{19}, 033102 (2012).
\bibitem{RPA low intensity} X. Zhang, B. Shen, X. Li, Z. Jin, and F. Wang, Phys. Plasmas \textbf{14}, 073101 (2007); O. Klimo, J. Psikal, J. Limpouch, V. T. Tikhonchuk, Phys. Rev. ST Accel. Beams \textbf{11}, 031301 (2008); A. P. L. Robinson, et al., New J. Phys.  \textbf{10}, 013021 (2008); B. Qiao, S. Kar, M. Geissler, P. Gibbon, M. Zepf, and M. Borghesi, Phys. Rev. Lett. \textbf{99}, 115002 (2012).
\bibitem{RPA exp} S. Kar, M. Borghesi, S. V. Bulanov, M. H. Key, T. V. Liseykina, A. Macchi, A. J. 
Mackinnon, P. K. Patel, L. Romagnani, A. Schiavi, O. Willi, Phys. Rev. Lett,  \textbf{100}, 225004 
(2008); K. U. Akli, S. B. Hansen, A. J. Kemp, R. R. Freeman, F. N. Beg, D. C. Clark, S. D. Chen, D. 
Hey, S. P. Hatchett, K. Highbarger, E. Giraldez, J. S. Green, G. Gregori, K. L. Lancaster, T. 
Ma, A. J. MacKinnon, P. Norreys, N. Patel, J. Pasley, C. Shearer, R. B. Stephens, C. Stoeckl, 
M. Storm, W. Theobald, L. D. Van Woerkom, R. Weber, M. H. Key, Phys. Rev. Lett. \textbf{100}, 
165002 (2008); A. Henig, S. Steinke, M. Schnurer, T. Sokollik, R. Horlein, D. Kiefer, D. Jung, 
J. Schreiber, B. M. Hegelich, X. Q. Yan, J. Meyer-ter-Vehn, T. Tajima, P. V. Nickles, W. 
Sandner, and D. Habs, Phys. Rev. Lett. \textbf{103}, 245003 (2009); C. A. J. Palmer, N. P. Dover, I. 
Pogorelsky, M. Babzien, G. I. Dudnikova, M. Ispiriyan, M. N. Polyanskiy, J. Schreiber, P. 
Shkolnikov, V. Yakimenko, and Z. Najmudin, Phys. Rev. Lett. \textbf{106}, 014801 (2011).
\bibitem{RT in RPA} F. Pegoraro and S. V. Bulanov, Phys. Rev. Lett. \textbf{99}, 065002 (2007).
\bibitem{unlimited} S. V. Bulanov, E. Yu. Echkina, T. Zh. Esirkepov, I. N. Inovenkov, M. Kando, F. Pegoraro, and G. Korn,
Phys. Rev. Lett. \textbf{104}, 135003 (2010); Phys. Plasmas \textbf{17}, 063102 (2010).
\bibitem{LightSail} A. Macchi, S. Veghini, and F. Pegoraro, Phys. Rev. Lett. \textbf{103}, 085003 (2009); A. Macchi, S. Veghini, T. V. Liseylina, and F. Pegoraro, New J. Phys. \textbf{12}, 045013 (2010).
\bibitem{MeV protons} T. Zh. Esirkepov, \textit{et al.}, JETP Lett. \textbf{70}, 82 (1999); 
A. M. Pukhov, Phys. Rev. Lett. \textbf{86}, 3562 (2001); 
Y. Sentoku, \textit{et al.}, Appl. Phys. B \textbf{74}, 207 (2002); 
A. J. Mackinnon, Y. Sentoku, P. K. Patel, D. W. Price, S. Hatchett, M. H. Key, C. Andersen, R. Snavely, and R. R. Freeman, Phys. Rev. Lett. \textbf{88}, 215006 (2002);
S. V. Bulanov, \textit{et al.}, JETP Lett. \textbf{71}, 407 (2000); 
Y. Sentoku, \textit{et al.}, Phys. Rev. E \textbf{62}, 7271 (2000); 
H. Ruhl, S. V. Bulanov, T. E. Cowan, T. V. Liseikina, P. Nickles, F. Pegoraro, M. Roth, W. Sandner, Plasma Phys. Rep.
\textbf{27}, 411 (2001).
\bibitem{CE} S. V. Bulanov, T. Zh. Esirkepov, V. S. Khoroshkov, A. V. Kuznetsov and F. Pegoraro, Phys. Lett. A
\textbf{299}, 240 (2002); E. Fourkal, I. Velchev, and C.-M. Ma, Phys. Rev. E \textbf{71}, 036421 (2005).
\bibitem{DCE} S. S. Bulanov, A. Brantov, V. Yu. Bychenkov, V. Chvykov, G. Kalinchenko, T. Matsuoka, P. Rousseau, S. Reed, V. Yanovsky, D. W. Litzenberg, and A. Maksimchuk, Med. Phys. \textbf{35}, 1770 (2008); 
S. S. Bulanov, A. Brantov, V. Yu. Bychenkov, V. Chvykov, G. Kalinchenko, T. Matsuoka, P. Rousseau, V. Yanovsky, D. W.
Litzenberg, K. Krushelnick, and A. Maksimchuk, Phys. Rev. E \textbf{78}, 026412 (2008).
\bibitem{REMP} T. Zh. Esirkepov, Comput. Phys. Comm. \textbf{135}, 144 (2001).
\bibitem{LI} S. V. Bulanov, \textit{et al.}, Phys. Rev. Lett. {\bf 91}, 085001 (2003); 
S. S. Bulanov,  \textit{et al.}, Phys. Rev. E {\bf 73}, 036408 (2006); 
M. Kando, \textit{et al.}, Phys. Rev. Lett. \textbf{99}, 135001 (2007); 
M. Kando, \textit{et al., ibid.} \textbf{103}, 235003 (2009).	
\bibitem{RM_ion} V. V. Kulagin, \textit{et al.}, Phys. Plasmas \textbf{14}, 	 (2007);
D. Habs, \textit{et al.}, Appl. Phys. B, {\bf 93}, 349 (2008);
T. Zh. Esirkepov, \textit{et al.},  Phys. Rev. Lett. \textbf{103}, 025002 (2009);
 H. Wu, \textit{et al., ibid.} \textbf{104}, 234801 (2010);
L. L. Ji, \textit{et al., ibid.} \textbf{105}, 025001 (2010).
S. S. Bulanov, \textit{et al.}, Phys. Lett. A, \textbf{374}, 476 (2010).
\bibitem{RT} G. Mourou, T. Tajima, and S. V. Bulanov, Rev. Mod. Phys \textbf{78}, 309 (2006).
\bibitem{RT exp} A. Henig, D. Kiefer, K. Markey, D. C. Gautier, K. A. Flippo, S. Letzring, R. P. Johnson, T. Shimada,
L. Yin, B. J. Albright, K. J. Bowers, J. C. Fernandez, S. G. Rykovanov, H.-C. Wu, M. Zepf, D. Jung, V. Kh. Liechtenstein,
J. Schreiber, D. Habs, and B. M. Hegelich, Phys. Rev. Lett \textbf{103}, 245003 (2009).
\bibitem{Shock} S. V. Bulanov, T. Zh. Esirkepov, M. Kando, F. Pegoraro, S. S. Bulanov,  C. G. R. Geddes, C. B. Schroeder, E. Esarey, W. Leemans, in preparation.
\bibitem{thin foil} V. A. Vshivkov, N. M. Naumova, F. Pegoraro, and S. V. Bulanov, Phys. Plasmas \textbf{5}, 2727 (1998).
\bibitem{optimal_esirkepov} T. Esirkepov, M. Yamagiwa, and T. Tajima, Phys. Rev. Lett. \textbf{96}, 105001 (2006). 
\end{thebibliography}
\end{document}